\documentclass[aps,prb]{revtex4}
\usepackage{graphicx}

\begin{document}
\pagestyle{empty}
\title{Frictional Drag induced in Low-
Dimensional Systems by Brownian Motion of Ions in Liquid Flow}

\author{A.I.Volokitin$^{1,2}$\footnote{Corresponding author.
\textit{E-mail address}:alevolokitin@yandex.ru}    and B.N.J.Persson$^1$}
 \affiliation{$^1$Institut f\"ur Festk\"orperforschung,
Forschungszentrum J\"ulich, D-52425, Germany} \affiliation{
$^2$Samara State Technical University, 443100 Samara, Russia}

\begin{abstract}
We study the  frictional drag force in low-dimensional systems (2D-electron  and
2D-liquid systems) mediated by a fluctuating electromagnetic field which originate from Brownian 
motion of ions in liquid. The analysis is focused on the [2D-system--2D-system], [2D-system
--semi-infinite liquid],  and [2D-system--infinite liquid] configurations. We show that for 2D-electron systems the friction drag depends linearly 
 on the relative velocity of the free carries of charge in the different media, but for 2D-liquid systems the frictional drag depends nonlinear  on the relative velocity.    
For  2D-systems the frictional drag force induced by liquid flow may be
several orders of magnitude larger than the frictional drag induced by an electronic current.
\end{abstract}

\maketitle


\vskip 5mm

\section{Introduction}
\begin{figure}
\includegraphics[width=0.45\textwidth]{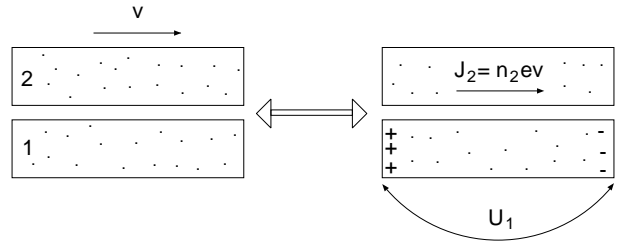}
\caption{\label{Fig1}Two way for studying noncontact friction. Left: Two bodies at finite separation are sliding relative to each other. The noncontact friction force between bodies is
 mediated by a fluctuating electromagnetic field. Right: 
A current of charged particles (electrons or ions) flow in upper medium. The fluctuating electromagnetic field associated with this motion will result in friction force 
acting on the free electrons or ions in medium at the bottom. This friction force will induce an electric field which can be measured.}
\end{figure}

All media are surrounded  by a fluctuating electromagnetic field 
because of the thermal and quantum fluctuations of the current density inside them.  The fluctuating field is responsible for many 
important phenomena such as radiative heat transfer, the van der Waals interaction, noncontact friction and frictional drag in low-dimensional systems 
\cite{Greffet6,RMP07}. The relation between fluctuations and friction is determined by the fluctuation-dissipation theorem. 
According to this theorem, the fluctuating force that makes a small particle jitter will also cause friction if the particle 
is dragged through the medium. Noncontact friction and frictional drag measurements are very closely related to each other.  In both these experiments 
the media are separated by a potential barrier thick enough to prevent electrons or other particle with a finite rest mass from tunneling 
across it, but allowing the interaction via the long-range electromagnetic field, which is always present in the gap between bodies.
In noncontact friction experiments the bodies  move relative to each other, while in frictional drag experiments the motion of charged particles (electrons 
or ions) is induced in one medium, and the reaction of  free carries of charge on this motion is measured in other medium 
(see Fig. \ref{Fig1})

There are two mechanisms of noncontact friction and frictional drag.  The \textit{electrostatic} friction is due to the relative motion of charged bodies. This mechanism of 
 noncontact friction was observed in \cite{Stipe,Kuehn2006}. In these 
experiments a charged probe tip oscillated close to the surface of substrate. Measurement of the damping of the probe oscillation 
 yields a noncontact friction coefficient that may be related to the spectrum of the electromagnetic field fluctuation at the probe frequency. Using the
fluctuation-dissipation theorem, the force fluctuations were
interpreted \cite{Stipe,Kuehn2006} in terms of near-surface fluctuating electric field
interacting with static surface charge. The electrostatic friction can be related to  the `image' charge which 
is induced in the sample by the charged tip. During motion of the tip relative to the sample the ``image'' charge will lag 
slightly behind the moving charge inducing it, and this is the origin of the electrostatic friction \cite{Volokitin2005,Volokitin2006}.   

Noncontact friction exists even between neutral bodies because the fluctuating electromagnetic field originated from charge fluctuations
in one medium will induce polarization in other medium. The interaction of the fluctuating electromagnetic field with the induced polarization is responsible
for many important phenomena  such as  radiative heat transfer and the van der Waals
interaction \cite{RMP07}.   When two media are in relative motion, the induced polarization will lag
behind the fluctuating polarization inducing it, and this gives rise  to the so-called  van der Waals friction.

\begin{figure}
\includegraphics[width=0.45\textwidth]{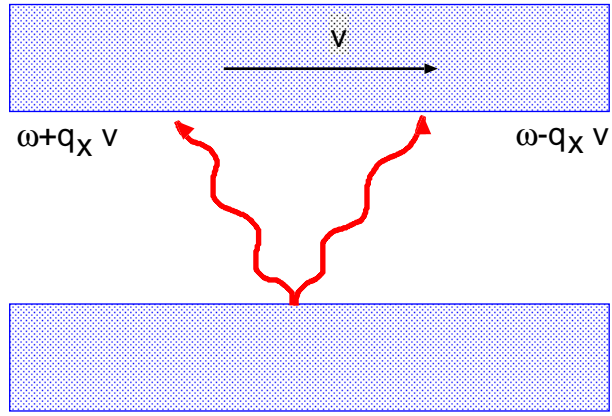}
\caption{\label{Fig2} Two bodies moving relative to each other
will experience van der Waals friction due to Doppler shift of the
electromagnetic waves emitted by them. }
\end{figure}

The origin of the van der Waals friction is  closely connected
with the Doppler effect.  Let us consider two flat parallel
surfaces, separated by a sufficiently wide insulator gap, which
prevents particles from tunneling across it. If the charge
carriers inside the volumes restricted by these surfaces are in
relative motion (velocity $v$) a frictional stress will act
between surfaces. This frictional stress is related to an
asymmetry of the reflection amplitude along the direction of
motion; see Fig. \ref{Fig2}.
If one body emits radiation, then in the rest reference frame of
the second body these waves are Doppler shifted which will result
in different reflection amplitudes. The same is true for radiation
emitted by the second body. The exchange of
``Doppler-shifted-photons'' will result in momentum transfer between bodies which
is  the origin of the  van der Waals friction.

\begin{figure}
\includegraphics[width=0.45\textwidth]{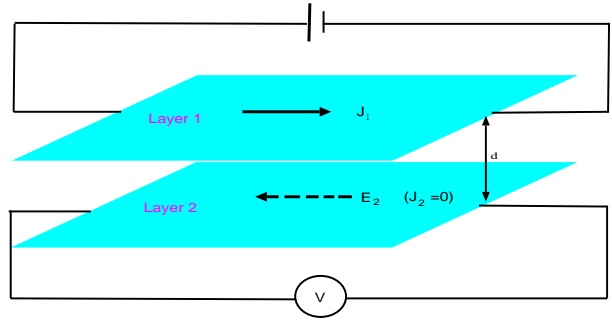}
\caption{\label{Fig3} A schematic diagram of a drag experiment. 
A current $\mathbf{J_1}$ is passed through layer \textbf{1}, and a voltmeter is attached to layer \textbf{2} to measure the induced electric field $\mathbf{E_2}$ 
due to interlayer interaction.} 
\end{figure}

The van der Waals friction can be probed not only by measuring the
friction force during relative motion of the two bodies, but an
alternative method consists in driving an electric current in one
metallic layer and studying of the effect of the frictional
drag of the electrons in a second (parallel) metallic layer (Fig.
\ref{Fig3}).
Such experiments were suggested by Pogrebinskii \cite{Pogrebinskii} 
and Price \cite{Price}, and were performed for 2D-quantum wells \cite{Gramila1,Sivan}. In these
experiments, two quantum wells are separated by a dielectric layer
thick enough to prevent electrons from tunneling across it but
allowing interlayer interaction between them. A current of density
$J_1=n_1ev$ is driven through layer \textbf{1} (where $ n_` $ is
the carrier concentration per unit area in the first layer), see
 Fig. \ref{Fig3}. Due to  the
interlayer interactions a frictional stress $\sigma =\gamma v$ will act on the
electrons in the layer \textbf{2} from layer \textbf{1} which will induce a current in layer \textbf{2}. If layer
\textbf{2} is an open circuit, an electric field $E_2$ will
develop in the layer whose influence cancels the frictional stress
$\sigma$ between the layers. Experiments \cite{Gramila1} show that, at least for small
separations, the frictional drag
 can be explained by the interaction between the electrons in the different layers via the fluctuating Coulomb field.
 However, for large
inter-layer separation
the friction drag is dominated by phonon exchange \cite{Bonsager}.

Recently,  it was observed that the flow  of a liquid over
bundles of single-walled carbon nanotubes (SWNT)
 induces a voltage in the sample along the  direction of the flow \cite{Kumar1,Kumar2}.
  The dependence of the voltage on the flow speed
was found to be  logarithmic  over five decades of variation of
the speed. There have been attempts to explain  this flow-induced
voltage in electrokinetic terms, as a result of the streaming
potential that develops along the flow of an electrolyte through a
microporous insulator \cite{Cohen,Ghosh}. Earlier Kr\'al and
Shapiro \cite{Kral} proposed that the  liquid flow transfer
momentum to the acoustic phonons of the nanotube, and that the
resulting  ``phonon wind'' drives  an  electric  current  in the
nanotube. They also suggested, qualitatively, that the fluctuating
Coulomb field of the ions in the liquid could drag directly the
carriers in the nanotube. However, the first mechanism \cite{Kral}
requires an enormous pressure  \cite{Kumar2}, while the second
mechanism \cite{Kral} result in a  very small current, of order
femtoAmperes \cite{Kumar2}. In \cite{Kumar2} another  mechanism
was proposed, which is related to the second idea of Ref.
\cite{Kral}, but which requires neither localization of carriers
nor drag at the same speed as the ions.   In fact, \cite{Kumar2}
considered the friction  between  a moving point charge and the
surrounding medium. This mechanism  of friction is described by the theory of electrostatic friction \cite{RMP07}. For neutral systems, such as a
nanotube, the \textit{electrostatic} friction proposed in
\cite{Kumar2} will  vanishing.

In \cite{Persson7} it was assumed that the liquid molecules nearest to the nanotube form a 2D-solidlike monolayer,
pinned to the nanotube
by  adsorbed ions. As the liquid flows, the adsorbed solid monolayer performs stick-slip type of sliding motion along the nanotube. The drifting adsorbed ions will
produce a voltage in the nanotube through electronic friction against free electrons inside the nanotube.

In \cite{Das} a model calculations of the frictional drag were presented involving  a channel containing
overdamped Brownian particles.
The channel  was embedded in a wide chamber containing the same type of Brownian particles with drift velocity
 parallel to the channel. It was found that the flow of particles in the chamber induces
a drift of the particles in the channel.

In this article we study 
frictional drag in low-dimensional system mediated by the fluctuating electromagnetic field originated from Brownian 
motion of ions. We compare this mechanism of frictional drag with the frictional drag resulting from charge fluctuations in electron system. 
A Brief Report about this work was published in \cite{Volokitin08}.

\section{van der Waals frictional drag between two 2D-systems}

 Let us consider two media with  flat parallel surfaces at separation $d\ll
\lambda_T=c\hbar/k_BT$. Assume that the free charge
carriers in one media move  with the velocity $v$ relative to other medium.
According to \cite{Volokitin6,Volokitin10} the frictional stress
between the two  media,     mediated by a fluctuating electromagnetic
field,  is determined by

\[
\sigma_{\|} =\frac \hbar {2\pi ^3}\int_{-\infty }^\infty dq_y\int_0^\infty
dq_xq_xe^{-2qd}\left\{ \int_0^\infty d\omega [n(\omega )-n(\omega
+q_xv)]\right.
\]
\[
\times \left( \frac{\mathrm{Im}R_{1p}(\omega +q_xv)\mathrm{Im}R_{2p}(\omega)
}{\mid 1-e^{-2
q d}R_{1p}(\omega +q_xv)R_{2p}(\omega)\mid ^2}+\left( 1\leftrightarrow 2\right) \right)
\]
\begin{equation}
\left. -\int_0^{q_xv}d\omega [n(\omega )+1/2]\left( \frac{\mathrm{Im}
R_{1p}(\omega -q_xv)\mathrm{Im}R_{2p}(\omega)}
{\mid 1-e^{-2qd}R_{1p}(\omega -q_xv)R_{2p}(\omega)\mid ^2}
+(1\leftrightarrow 2)\right) \right\}.
\label{parallel2}
\end{equation}
where  $n(\omega )=[\exp (\hbar \omega /k_BT-1]^{-1}$ and
($1\leftrightarrow 2$) denotes the terms which are obtained from
the preceding  terms
 by permutation of indexes
$1$ and $2$. $R_{ip}$ ($i=1,2$) is the reflection amplitude for
surface $i$ for $p$ -polarized electromagnetic waves. The
reflection amplitude for a 2D- system is determined by
\cite{Volokitin10}
\begin{equation}
R_{ip}=\frac{\epsilon _{ip}-1}{\epsilon _{ip}+1},
 \label{refcoef}
\end{equation}
where $\epsilon _{ip}= 4\pi iq \sigma _i(\omega,q)/\omega
\varepsilon +1$, $\sigma _i$  is the longitudinal conductivity
of the layer $i$, and $\varepsilon$ is the dielectric constant of
the surrounding dielectric.

For a 2D-electron system the longitudinal conductivity can be
written in the form $\sigma _l(\omega ,q)=-\mathrm{i}\omega \chi
_l(\omega ,q)/q^2,$ where $\chi _l$ is the finite life-time
generalization of the longitudinal Lindhard response function for
a 2D-electron gas \cite{Mermin,Stern}
\begin{equation}
\chi _l(\omega ,Q)=\frac{(1+i/\omega \tau )\chi _l^0(\omega
+i/\tau ,Q)}{ 1+(i/\omega \tau )\chi _l^0(\omega +i/\tau ,Q)/\chi
_l^0(0,Q)}, \label{rthree}
\end{equation}
where for a degenerate electron gas
\begin{equation}
\chi _l^0(\omega ,q)=\frac{n_se^2}{zm^{*}v_F^2}\left\{
2z+\sqrt{(u-z)^2-1}- \sqrt{(u+z)^2-1}\right\} ,\label{Lindhard}
\end{equation}
where $\tau $ is the relaxation time, $z=q/2k_F$, $u=\omega
/(qv_F)$, and $k_F$ and $v_F$ are the Fermi wave vector and Fermi
velocity, respectively, and $n_s$ is the 2D-electron density in the
layer.
 We define an effective electric field equal to the friction force per unit charge:
$E=\sigma_{\|}/n_se$. For $v\ll v_F$, where $v_F$ is the  Fermi
velocity, the friction force  depends linearly on
velocity $v$. For $d=175\ \mathrm{\AA }$ at $T=3$ K, and with
$n_s=1.5\times 10^{15}$ m$^{-2}$, the electron effective mass
$m^{*}=0.067$ m$_e$,
$v_F=1.6\times 10^7$ cm/s,   the electron mean free path
$l=v_F\tau =1.21\times 10^5\ \mathrm{\AA }$, and $\varepsilon =10$ (which
corresponds to the condition of the experiment \cite{Gramila1}) we get  $E=6.5\times
10^{-6}v$ V/m, where the velocity $v$ is in m/s. For a current $200$ nA in
a two-dimensional layer with the width $w=20\mu$m the drift of electrons
 (drift velocity $v=60$m/s) creates a frictional electric field in
the adjacent quantum well $E=4\cdot10^{-4}$V/m. Note that for the electron systems the frictional drag force decreases
when the electron concentration increases. As an
a example, for 2D-quantum wells with high electron
density ($n_s=10^{19}$ m$^{-2}$, $T=273$ K, $\tau =4\times
10^{-14}$ s,  $\varepsilon =10$, $m^*=m_e$) at $d=175\ \mathrm{\AA }$ we get
$E=1.2\times 10^{-9}v$ V/m.the thickness of the channel $d_c=10$ nm

Let us now consider  a fluid  with the ions in a narrow channel with thickness $d$.
For $d\gg q^{-1}_{D}\gg d_c$, where $q_{D}=\sqrt{4\pi
N_0Q^2/\varepsilon_c k_BT}$ is the Debye screening wave number ($N_0$ is the  concentration of ions, and $\varepsilon_c$ is the
dielectric constant of the liquid in the channel, and  $Q$ is the ion charge), the channel can be considered as
two-dimensional. The Fourier transform of the
 diffusion equation for the ions (of type \textit{a}) in the channel can be written in
the form
\begin{equation}
\frac{i\omega}{D_a}\sigma^a_q = q^2\left(\sigma^a_q + \frac
{N_aQ^2d_c}{k_BT}\varphi_q\right), \label{refcoef11}
\end{equation}
where $\sigma^a_q$ and $\phi_q$ are the Fourier components of the
surface charge density  and the electric potential, respectively, and $D_a$ is the  diffusion coefficient of the ions in the liquid in
the channel. From Eq. (\ref{refcoef11}) we get
\begin{equation}
\sigma^a_q = -\frac
{N_aQ^2d_c}{k_BT}q^2\frac{\varphi_q}{q^2-i\omega/D_a}.
\label{refcoef12}
\end{equation}
The surface current density resulting from the diffusion and drift of
the ions of type \textit{a}, is determined by the formula
\[
j^a_{iq} = -iqD\left(\sigma^a_q +
{N_aQ^2d_c}{k_BT}\varphi_q\right)=
\]
\begin{equation}
= -i\omega \frac {N_aQ^2d_c}{k_BT}\frac{1}{q^2-i\omega/D_a}E_q,
\label{refcoef13}
\end{equation}
where $E_q=-iq\varphi_q$ is the Fourier component of the electric
field. Furthermore, there is a surface current density connected
with the polarization of the liquid, which is determined by the
formula
\begin{equation}
j_{pq} = -i\omega p_q = -i\omega d_c \frac {\varepsilon_ñ
-1}{4\pi}E_q, \label{refcoef13}
\end{equation}
where $p_q$ and $\varepsilon_c$ are the surface polarization and
dielectric permeability of liquid in the channel, respectively.
Thus the total current density $j_q=
\sigma (\omega, q) E_q$, where the conductivity of the 2D-liquid   is determined by the
formula
\begin{equation}
\sigma(\omega,q) = -\frac{i\omega
d_c}{4\pi}\left(-1+\varepsilon_c\left(1+\sum_a\frac{q_{D}^2}{q^2-i\omega/D_a}\right)\right).
\label{conductivity}
\end{equation}

\begin{figure}
\includegraphics[width=0.45\textwidth]{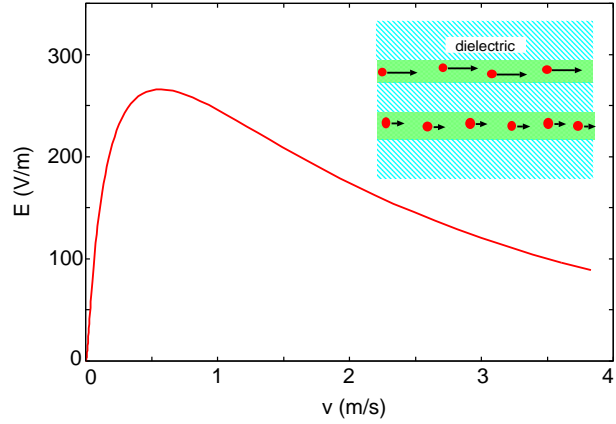}
\caption{\label{Fig4} The effective electric field  in a 2D-channel
with liquid as a function of the  flow velocity           
 in second 2D-channel  for
identical liquids in both channels. The temperature $T=300$ K, the ion concentration in
liquid $N_0=10^{24}$m$^{-3}$, the thickness of the channels $d_c=10$ nm,
 the diffusion coefficients of ions $D=10^{-9}$m$^2$/s and the dielectric
constant of the liquid $\varepsilon_c=80$. The dielectric
constant of the dielectric in the gap between channel $\varepsilon=10$ and  the separation between the channels  $d=175$ \AA.}
\end{figure}

For the [2D-electron]--[2D-liquid] configuration with the same parameters as above for electron system (with high 
electron density) at $d=175$ {\AA}, and with $\varepsilon_c=80$, $N_0=10^{24}$m$^{-3}$, $D=10^{-9}$m$^2$/s, $d_c=100$ {\AA} 
we get $E=4.6\times 10^{-8}v$ V/m, which is one order of magnitude 
larger than for the [2D-electron]--[2D-electron] configuration (with high electron density). Fig. \ref{Fig4} shows the dependence of the effective electric field in 
the channel on velocity of the liquid flow in ajancent channel, with the same  liquid.  
For comparison with [2D-electron]--[2D-electron] and [2D-electron]--[2D-liquid] configurations, for the [2D-liquid]--[2D-liquid]
 configuration the effective electric field is many orders of 
magnitude larger, and depends nonlinerly  on the liquid flow velocity $v$.

\section{van der Waals frictional drag in a 2D- system
induced by liquid flow in a semi-infinite chamber}

 Let us consider a 2D-electron system, isolated from a semi-infinite
liquid flow by a dielectric layer with the thickness  $d$. For the
2D- electron system the reflection amplitude is determined by Eqs. (\ref{refcoef}) - (\ref{Lindhard}). To find  the reflection amplitude   for
interface between the dielectric and the liquid we will assume
that the liquid fills half-space  $z \geq 0$, and that the half-space
with $z< 0$  is filled by a dielectric with the dielectric constant
$\varepsilon$. Let us study  the reflection of
an electromagnetic wave from the surface of the liquid in the nonretarded
limit, which formally corresponds to the limit $c\rightarrow\infty$. In
the region  $z< 0$ the potential can be written in the form
\begin{equation}
\varphi_q = (e^{-qz} - Re^{qz})e^{i\mathbf{q\cdot x}-i\omega t},
\label{1}
\end{equation}
where  $q$ is the magnitude of the component of the wave vector parallel to surface. We
will assume that the liquid consists of  ions of two types 
$a$ and $b$. The 
equation of continuity for the ions
\begin{equation}
-i\omega n_i + \mathbf{\nabla}\cdot \mathbf{J}_i = 0, \label{2}
\end{equation}
where $i = a, b$, $n_i = N_i - N_0$, where $N_i$ and $N_0$  are the
concentration of ions in the presence and absence of the electric field,
respectively. To linear order in  the electric
field
\begin{equation}
\mathbf{J}_i = -N_0\mu_iQ_i\mathbf{\nabla}\varphi -
D_i\mathbf{\nabla}n_i, \label{3}
\end{equation}
where $D_i$ is the diffusion coefficient, $\mu_i$ is the mobility
and $Q_i$ is the charge for ions of type $i$.
The diffusion coefficient and the mobility  are related with each other
by the Einstein relation: $D_i = k_BT\mu_i$. We  consider the case
when the  different ion mobilities  differ  considerably. In this case, in the calculation of dielectric response it
is possible to disregard the diffusion of the less mobile ions.
Omitting the index  $i$ for the more mobile ions, after 
substitution of  (\ref {3}) in  (\ref {2}) we obtain 
\begin{equation}
i\omega n + D\nabla^2 \left(n+ \frac{N_0Q}{k_BT}\varphi\right) = 0.
\label{4}
\end{equation}
This equation must be supplemented with Poisson's equation
\begin{equation}
\nabla^2 \varphi = - \frac{4\pi Qn}{\varepsilon_0}, \label{5}
\end{equation}
where $\varepsilon_0$ is the  dielectric permeability of the liquid. The
general solution of equations (\ref {4}) and (\ref {5}) can be
written in the form
\begin{equation}
\varphi = (C_1 e^{-\lambda z} + C_2 e^{-q z})e^{i\mathbf{q\cdot
x}}, \label{lref5}
\end{equation}
where $\lambda = \sqrt{q^2 + q_D^2 -i\omega/D}$ and  $q_D = \sqrt{4\pi
N_0Q^2/\varepsilon_0k_BT}$. At the interface  ($z=0$) the electric
potential and the normal   component of the electric
displacement field  must  be continuous, and the normal component
 of the flow density  must vanish. From these
boundary conditions we obtain 
\begin{equation}
C_1 + C_2 = 1 - R, \label{6}
\end{equation}
\begin{equation}
-\varepsilon_0 (\gamma C_1 + qC_2) + \varepsilon q(1+R) =0,
\label{7}
\end{equation}
\begin{equation}
i\omega\gamma C_1 + Dq_D^2qC_2=0. \label{8}
\end{equation}
From  (\ref{6})-(\ref{8}) we get
\begin{equation}
R = \frac{\epsilon -1}{\epsilon +1}, \label{9}
\end{equation}
where
\begin{equation}
\epsilon=\frac{\varepsilon_0\lambda
(Dq_D^2-i\omega)}{\varepsilon(Dq_D^2q-i\omega\lambda)}.
\label{refcoef10}
\end{equation}
For $v \ll v_F$ the frictional drag force acting on  the electrons
in  the 2D- system, due to the interaction with the ions in the liquid,
increases linearly with  the fluid velocity  $v$. In particular,
for $N_0=10^ {24} $m$^ {- 3}$, $T=273$ K, $\varepsilon_0=80$,
$D=10^ {- 9} $m$^2$/s, for a high electron density  ($n_s=10^
{19} $m$^ {- 2}$) in  the 2D- electron system, $E=1.4 \cdot 10^ {-
6} v$ V/m. This  effective electric field is three orders of magnitude larger than
obtained for  two 2D- electron systems with  high electron
concentration, and of the same order of magnitude as friction
between two  2D- electron systems with  low electron
concentration.

\begin{figure}
\includegraphics[width=0.45\textwidth]{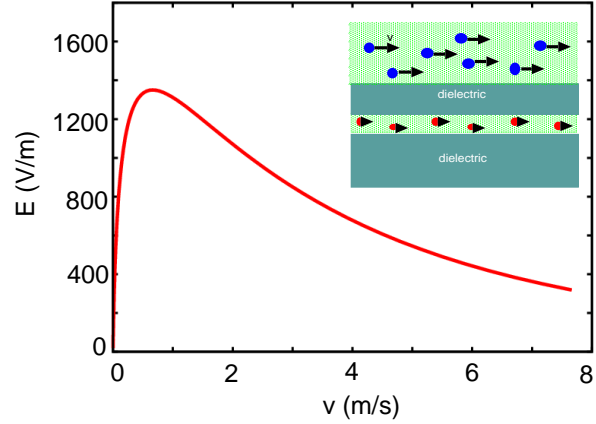}
\caption{\label{Fig5} The effective electric field  in a 2D-channel
with liquid induced by liquid flow 
 in a semi-infinite chamber as a function of the  flow velocity for
identical liquids in the channel and in the chamber. The temperature $T=300$ K, the ion concentration in
liquid $N_0=10^{24}$m$^{-3}$, the thickness of the channel $d_c=10$ nm,
 the diffusion coefficients of ions $D=10^{-9}$m$^2$/s and the dielectric
constant of the liquid $\varepsilon_0=80$. The dielectric
constant of the dielectric in the gap between channel $\varepsilon=10$ and 
the separation between the channel and semi-infinite chamber  $d=1$ nm.}
\end{figure}

 Fig. \ref{Fig5}  shows the dependence of the effective electric field 
in the 2D- channel  on the   velocity of the liquid flow  in the
semi-infinite chamber, for identical liquid in the channel and in
the  chamber. We have used the same parameters as  above for
the liquid,  with the separation between the channel and chamber
$d=1$nm. The effective electric field  in the channel
initially increases with the flow velocity, reaches a maximum, and
then decreases, in agreement
with the model calculation in \cite{Das}. The position of the maximum
decreases when the density of ions decreases. The frictional drag
force induced by the liquid flow in the narrow channel is 9 orders
of magnitude larger than the frictional drag  force  induced in a
2D-electron system.

\section{van der Waals frictional drag in a 2D- system
induced by   liquid  flow in an infinite chamber}

As a limiting case of the situation considered above, let us
consider a 2D-system immersed in a flowing liquid in an infinite
chamber. We assume that the liquid flows  along the $x$-axis, and
that the plane of 2D-system coincides with the $xy$- plane.
According to the fluctuation-dissipation theorem, for an infinite medium  the correlation function for the Fourier components of
the longitudinal current density   is
determined by \cite{RMP07}
\begin{equation}
<j_k^fj_k^{f*}>_{\omega} = \frac{\hbar}{(2\pi)^2}[n(\omega)
+1/2]\omega^2\mathrm{Im}\varepsilon (\omega , k), \label{Inl1}
\end{equation}
where  $\mathbf{k}=(\mathbf{q},k_z)$ is the wave vector. The longitudinal current density
is connected with the charge density $\rho_k =  kj_k /\omega$ via the continuity equation, from
which one obtain
\begin{equation}
<\rho_k^f\rho_k^{f*}>_{\omega} = \frac{\hbar}{(2\pi)^2}[n(\omega)
+1/2]k^2\mathrm{Im}\varepsilon (\omega , k). \label{Inl2}
\end{equation}
Poisson's equation for the electric potential gives
\begin{equation}
\varphi_k^f = \frac{4\pi \rho_k}{k^2\varepsilon (\omega , k)}.
\label{Inl3}
\end{equation}
In the  $xy$-plane the \textbf {q} -  component of the electric potential is
determined by
\begin{equation}
\varphi^f(\omega,q,0) = \int\frac {dk_ze^{ik_z0^+}}{2\pi}\varphi_k^f.
\end{equation}
From  (\ref{Inl2}) and (\ref{Inl3}) we get
\begin{equation}
<\varphi^f (\omega,q,0)\varphi^{*f} (\omega,q,0)>=4\hbar
(n(\omega) +1/2)\mathrm{Im}\Sigma(\omega,q,0),\label{one}
\end{equation}
where
\begin{equation}
\Sigma(\omega,q,0) = -\int_{-\infty}^{\infty}\frac{dk_z}{2\pi}\frac{1}{k^2\varepsilon_0\varepsilon(\omega,k)}.
\label{Inlsigma}
\end{equation}
Taking into account  that $E_q = iq \varphi_q$ we get $<E^f_q
E^{*f}_q>_{\omega}=q^2<\varphi^f_q \varphi^{*f}_q>_{\omega}$. From
the   diffusion and Poisson's equations we get
\begin{equation}
\frac{i\omega}{D} \rho_k = k^2\left(\rho_k + \frac
{N_0Q^2}{k_BT}\varphi_k\right) ,\label{Inl14}
\end{equation}
\begin{equation}
\varepsilon_0k^2\varphi_k = 4\pi\rho_k + 4\pi\rho_k^f.
\label{Inl15}
\end{equation}
From  (\ref{Inl14}) and (\ref{Inl15}) we get
 the dielectric function of the Debye plasma
\begin{equation}
\varepsilon(\omega,k)=1+\frac{q_D^2}{k^2-i\omega/D}. \label{Inleps}
\end{equation}
Substituting (\ref{Inleps}) in (\ref{Inlsigma}) gives
\begin{equation}
\Sigma(\omega,q)=-\frac{1}{\varepsilon_0(q_D^2-i\omega/D)}\left[-\frac{i\omega/D}{2q}+\frac{q_D^2}{2\gamma}\right].
\label{Inlsigma1}
\end{equation}
 According to the fluctuation-dissipation theorem,
the average value of the correlation function for the Fourier
components of the fluctuating surface charge density in the 2D-system
is determined by \cite{RMP07}
\begin{equation}
<\tau_q^f\tau_q^{f*}>_{\omega}=\frac{\hbar q^2}{\pi
\omega}(n(\omega) +1/2) \mathrm{Re}\sigma (\omega,q).
\label{1Inl20}
\end{equation}

If the 2D- system is surrounded by  liquid flow the electric
field created by the fluctuations of the charge density in the
fluid will induce  surface charge density fluctuations in the 2D- system. The spectral correlation 
functions (\ref {Inl14}) and
(\ref {1Inl20}) are determined in the rest reference frame of the
 liquid, and of the  2D-system, respectively.
In order to find the connection between the electric fields in
different reference frames    we  use the Galileo
transformation which leads to the Doppler frequency shift of the
electrical field  in the different reference frames. The electric field
in the plane of the 2D- system, due to the fluctuations of the charge
density in the liquid, will take the form
\begin{equation}
E(\mathbf{x},t) = e^{-i(\omega +q_xV)t +i\mathbf{q\cdot
x}}E_q^{I}, \label{Inl16}
\end{equation}
where $E_q^I$ is  the sum of the electric fields
created by the fluctuations of the charge density in the fluid and
the induced charge density in the 2D- system:
\begin{equation}
E_q^{I} = E_q^f + 4\pi
iq\Sigma(\omega,q)\tau^{I}_q(\omega^+), \label{Inl17}
\end{equation}
where $\omega^+ = \omega +q_xV$ and  $\tau^{I}_q$ is the surface induced
charge density. According to Ohm's law
\begin{equation}
j_q^{I} = \sigma_q^+E_q^I=\sigma_q^+(E_q^f + 4\pi
iq\Sigma(\omega,q)\tau^{I}_q(\omega^+)), \label{Inl18}
\end{equation}
where $\sigma_q^+=\sigma(\omega^+,q)$ is the  longitudinal conductivity
for the 2D-system. The
continuity equation   for the surface charge density gives
$j_q^{ind}=\omega^+\tau^{ind}_q/q$ and from  (\ref{Inl18}) we get
\begin{equation}
\tau_q^{I} = \frac{q}{\omega^+}\frac
{\sigma_q^+E_q^f}{1-4\pi i q^2\sigma_q^+/\omega^+\Sigma(\omega,q)}
\label{Inl19}
\end{equation}
and
\begin{equation}
E^{I}_q = \frac {E_q^f}{1-4\pi
iq^2\sigma_q^+/\omega^+\Sigma(\omega,q)}. \label{Inl20}
\end{equation}

In order to find the electric field created by the  charge
density fluctuations in the 2D- system it is necessary to solve
Poisson's equation in the rest reference frame  of the liquid. In this
reference frame the  charge density takes the form
\begin{equation}
\tau(\mathbf{x},t) = e^{-i(\omega -q_xV)t +i\mathbf{q\cdot
x}}\tau_q^{II}. \label{Inl16}
\end{equation}
Note that  the charge density  is composed from the fluctuating
$\tau^f$ and induced  $\tau^{ind}$ charge density: $\tau_q^{II}=\tau_q^f + \tau_q^{ind}$. In the
presence of the liquid flow   the  electric field in the plane of
the 2D- system, due to the fluctuating  surface charge density,  is
determined by
\begin{equation}
E_q^{II}=4\pi iq\Sigma(\omega^-,q)(\tau_{q}^{ind} +
\tau_{q}^f), \label{Inl21}
\end{equation}
where $\omega^- = \omega -q_xV$. From Ohm's law we get the following expression
for the induced charge density
\begin{equation}
\tau_{q}^{ind} = \frac{4\pi
iq^2\sigma(\omega,q)\Sigma(\omega^-,q)}{\omega}\frac{\tau^f}{1-4\pi
iq^2\sigma_q\Sigma(\omega^-,q)/\omega}. \label{Inl22}
\end{equation}
Substituting (\ref{Inl22}) in (\ref{Inl21}) we get
\begin{equation}
E_q^{II}=\frac{4\pi iq\Sigma(\omega^-,q)\tau^f}{1-4\pi
iq^2\sigma_q\Sigma(\omega^-,q)/\omega} \label{Inl23}
\end{equation}
and
\begin{equation}
\tau_q^{II}=\frac{\tau^f}{1-4\pi
iq^2\sigma_q\Sigma(\omega^-,q)/\omega}. \label{Inl24}
\end{equation}

\begin{figure}
\includegraphics[width=0.45\textwidth]{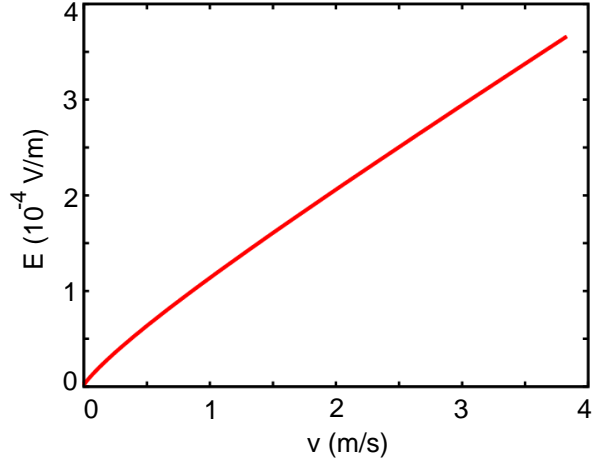}
\caption{\label{Fig6} The effective electric field  in a 1D-electron system
 induced by liquid flow
 in a infinite chamber, as a function of the  flow velocity. For the same parameters for the liquid as in Fig. \ref{Fig5}. 
The electron
concentration per unit length in 1D-system $n_l=3\cdot10^9$m$^{-1}$ and  the electron relaxation
time $\tau = 4\cdot 10^{-14}$s}
\end{figure}

The friction force per unit area of the 2D-system is given by
\begin{equation}
\sigma_{\|}=\int^{\infty}_{-\infty} d\omega \int
\frac{d^2\mathbf{q}}{(2\pi)^2}\frac{q_x}{q}<
E_q\tau_q^*>_{\omega}, \label{Inl25}
\end{equation}
where $E_q = E_q^{I}+E_q^{II}$ and  $\tau=\tau^I + \tau^{II}$.
Substituting (\ref{Inl19}), (\ref{Inl20}) and (\ref{Inl23}),
(\ref{Inl24}) in (\ref{Inl25}) we get
\[
\sigma_{\|} =\frac {2\hbar} {\pi ^2}\int_{-\infty }^\infty
dq_y\int_0^\infty dq_xq_xq^2\Big\{ \int_0^\infty d\omega [n(\omega
)-n(\omega +q_xv)]
\]
\[
\times \left(
\frac{\mathrm{Re}\sigma(\omega+q_xv)\mathrm{Im}\Sigma(\omega,q)
}{(\omega+q_xv)\mid 1-4\pi iq^2\sigma(\omega
+q_xv)\Sigma(\omega,q)/(\omega +q_xv)\mid ^2}+\left( \omega +q_xv
\leftrightarrow \omega\right) \right)
\]
\[
 -\int_0^{q_xv}d\omega [n(\omega )+1/2] \Big( \frac{\mathrm{Re}\sigma(\omega-q_xv)\mathrm{Im}\Sigma(\omega,q)}
{(\omega-q_xv)\mid 1-4\pi iq^2\sigma(\omega
-q_xv)\Sigma(\omega,q)/(\omega -q_xv)\mid ^2}
\]
\begin{equation}
 +( \omega -q_xv
\leftrightarrow \omega)\Big) \Big\}, \label{Inl26}
\end{equation}
where ($ \omega \pm q_xv \leftrightarrow \omega$)  denote the
terms which are obtained from the preceding  terms by permutations
of the   arguments $\omega \pm q_xv$
 and $\omega$. With the same parameters as used above for the liquid, and for the high density 2D-electron
 system, we get
$E=8.1\cdot 10^{-6}v$ V/m. For a 1D-electron system we obtained a formula which is similar to Eq. (\ref{Inl26}).
Fig. \ref{Fig6}  shows the result
of the calculations of the effective electric field for a 1D-electron system with the electron density per unit length $n_l=
3\times 10^9$m$^{-1}$, the temperature $T=300$ K, and with the same parameters for the liquid as used above.
For the 1D-electron system we obtained a
slight deviation
from the linear dependence of the frictional drag on the  liquid flow velocity. The frictional drag for the
1D-electron system is one order
of magnitude larger than for the 2D-electron system.

\begin{figure}
\includegraphics[width=0.45\textwidth]{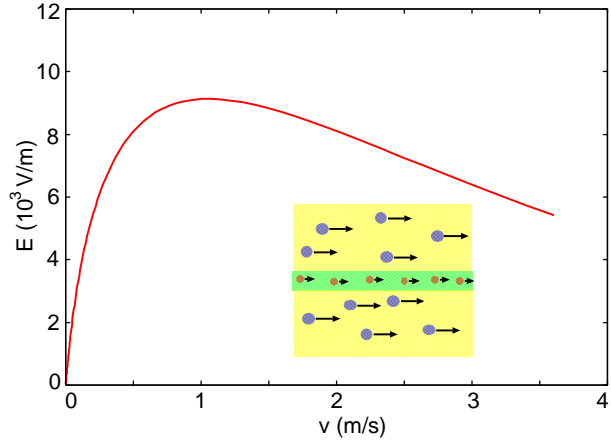}
\caption{\label{Fig7} The same as  Fig. \ref{Fig5} but for infinite chamber.}
\end{figure}

Fig. \ref{Fig7}  shows
the dependence of the effective electric field  in the liquid
in the 2D-channel  on the
liquid flow velocity in the infinite chamber,
assuming identical liquid in the channel and in the
chamber. Qualitatively, we obtained the same results for a 1D-channel.

\section{Discussion and conclusion}

For a channel with open ends  the frictional drag force will
induce a drift motion of the ions in the liquid with the velocity
$v_d=D_cQE/k_BT$. The positive and negative ions will drift in the
same direction. If ions have different mobility then  the drifting
ions  will lead to an electric current whose direction will be
determined by the current created by the ions with the largest  mobility.
For a channel with  closed ends the frictional drag force
 will lead to a change in ion concentration along the
channel. In the case of ions with the different mobilities,
the friction force will be  different for the ions with the
opposite charges. As a result the ions of opposite charges will
be characterized by different distribution functions which, as for electronic systems, will result in an electric field, 
and  an induced voltage which can be measured. Let us write the friction force acting on the ions of different 
type in the form: $F_a=QE_a$ and $F_b=QE_b$. From the condition that, in the static case, the flux density in the channel must vanish, we get
\begin{equation}
n_a = -\frac {Q}{k_BT}\left (\varphi -E_ax\right ), \label{con1}
\end{equation}
\begin{equation}
n_b = \frac {Q}{k_BT}\left (\varphi +E_ax\right ). \label{con2}
\end{equation}
These equation must be supplemented with Poisson's equation
\begin{equation}
\frac {d^2\varphi}{d^2x}=-\frac{4\pi Q}{\varepsilon_c}(n_a - n_b). \label{con3}
\end{equation}
Substituting  (\ref{con1})-(\ref{con2}) in  (\ref{con3}) we get
\begin{equation}
\frac {d^2\varphi}{d^2x}=q_D^2\left (2\varphi - \Delta Ex\right), \label{con4}
\end{equation} 
where $\Delta E = E_a - E_b$. The solution of  (\ref{con4}) with boundary condition 
\begin{equation}
\frac {d\varphi}{dx}\Big|_{x=\pm L/2}=0,
\end{equation}
where $L$ is the channel length, has the form
\begin{equation}
\varphi(x) = \frac {\Delta E}{2}\left (x - \frac {1}{\sqrt{2}q_D}\frac {\sinh{\sqrt{2}q_Dx}}
{\cosh \sqrt{2}q_DL/2}\right). \label{con5}
\end{equation}
The voltage between the ends of the channel is determined by
\begin{equation}
U = \varphi (L/2) - \varphi (-L/2) = \Delta E\left (\frac {L}{2} - \frac {1}{\sqrt{2}q_D}\tanh {\sqrt{2}q_DL/2}\right). \label{con6}
\end{equation}
For $q_DL \gg 1$
 the voltage, which
appears as a result of the frictional drag, will be approximately equal
to  $U \approx \Delta{E}L/2$. Furthermore, the
frictional drag will induce a pressure difference $\Delta p=nLeE$.
For example, if $N_0=10^{24}$m$^{-3}$, $L=100\mu$m and $E=1000$V/m
we get pressure difference $\Delta p=10^4$ Pa, which should be
easy to measure. Assume now that  one type of ions are fixed (adsorbed) on the walls of the channel and
an  equal number of mobile ions of opposite sign are distributed
in the liquid phase. In this case the motion of the polar liquid in
the adjacent region will lead to  frictional drag force acting
on the mobile ions in the channel. For a channel with the closed
ends this frictional drag will induce a
 voltage, which can be measured.

In this paper we have shown that the van der Waals frictional drag
force, induced in low-dimensional system by liquid flow, can be
several orders of magnitude larger than the friction induced by an
electron current. For  narrow 2D-and 1D-channels with liquid the
frictional drag force is several orders of magnitude larger than
for 2D-and 1D-electron systems. In the contrast to electron systems,
the frictional drag force for a narrow channel with liquid depends
nonlinearly  on the flow velocity. These results contradict to the
calculations in Ref. \cite{Kumar2}, where it was assumed that the
observed nonlinear dependence of a voltage on the liquid flow
velocity   was connected with the frictional drag   acting on the electrons
in the nanotubes.  However, according to our calculations,  the
mobile ions in the channels of porous medium experience considerably
greater frictional drag force  due to van der Waals friction, than
on the electrons in low-dimensional electronic structures. Thus,
the liquid flow induced voltage observed in \cite{Kumar2}  is more
likely connected with the frictional drag experienced by  mobile ions in an
electrolyte in the channels between nanotubes, rather than the  electrons
in the nanotubes.  These results should have a broad
application for studying of the van der Waals friction and in the
design of nanosensors. Such  detectors would be of
great interest in micromechanical and biological applications
\cite{Schasfoort,Munro}, where local dynamical effects are
intensively studied.

\vskip 0.5cm \textbf{Acknowledgment }

A.I.V acknowledges financial support from the Russian Foundation
for Basic Research (Grant N 08-02-00141-a) and DFG.

\end{document}